\documentstyle[]{article}
\newcommand{\heading}[1]{\noindent{\Large\bf{#1}}}
\renewenvironment{abstract}{\vskip 2mm
\noindent {\bf Abstract.\,}
  \small
}{}
\newcounter{adrc}
\renewcommand{\author}[1]{\vskip 1mm \noindent{\sc  #1}\vskip 1mm}
\newcommand{\address}[1]{\noindent$^{\theadrc}${\small\it #1}
\addtocounter{adrc}{1}\\}
\def\lsim{~\rlap{$<$}{\lower 1.0ex\hbox{$\sim$}}}
\def\gsim{~\rlap{$>$}{\lower 1.0ex\hbox{$\sim$}}}
\newenvironment{iapbib}[1]{\footnotesize
\begin{thebibliography}{#1}
\setlength{\itemsep}{-0.5mm}}
{\end{thebibliography}}
\begin{document}
\heading{Non-Markovian effects in the solar neutrino problem}
\par\medskip\noindent
\author{
        G. Gervino$^{1,2}$,
        G. Kaniadakis$^{2,3}$,
        A. Lavagno$^{2,3}$,   
        M. Lissia$^{4,5}$ and
        P. Quarati$^{3,4}$
       }
\address{Dipartimento di Fisica, Universit\`a di Torino, I-10125 Torino, Italy}
\address{Istituto Nazionale di Fisica Nucleare, Sezione di Torino,
         I-10125 Torino, Italy}
\address{Dipartimento di Fisica and INFM, Politecnico di Torino,
         I-10129 Torino, Italy}
\address{Istituto Nazionale di Fisica Nucleare, Sezione di Cagliari,
         I-09042 Monserrato, Italy }
\address{Dipartimento di Fisica, Universit\`a di Cagliari,
         I-09042 Monserrato, Italy }
\begin{abstract}
The solar core, because of its density and temperature, is not a 
weakly-interacting or a high-temperature plasma. Collective effects have 
time scales comparable to the average time between collisions, and the
microfield distribution influences the particle dynamics.
In this conditions ion and electron diffusion is a non-Markovian process,
memory  effects are present and the equilibrium statistical distribution
function differs from the Maxwellian one. 
We show that, even if the deviations from the standard velocity distribution
that are compatible with our present knowledge of the solar interior
are small, they are sufficient to sensibly modify the sub-barrier nuclear
reaction rates. The consequent changes of the neutrino fluxes are
comparable to the flux deficits that constitute the solar neutrino
problem.
\end{abstract}

\vspace{0.5cm}

The solar neutrino problem is one of the most interesting long standing puzzle 
of the modern physics. The discrepancy between the combined results from the 
solar neutrino experiments (Homestake, GALLEX, SAGE, Kamiokande and
SuperKamiokande) and the predictions of the standard solar models (SSM)
combined to the minimal standard electroweak model has suggested the
hypothesis that neutrinos could have small masses 
and the lepton flavor is not conserved~\cite{bahcall,caste}. 
Because the neutrino oscillation theory has far-reaching consequences for both 
particle physics and cosmology, it is of great importance to ask whether
the solar  neutrino problem can be solved, or at least alleviated, in the
framework of the conventional physics~\cite{dar}. 

The aim of this contribution is to show that the physical characteristics
of the solar core (the plasma parameter $\Gamma=(Z e)^2 / (a k T) $,
where $a$ is the interparticle average distance, the collisional frequency
$\nu_c$ and the electric microfield intensity and distribution) indicate
the presence of long-range many-body interactions and memory effects.
Therefore, the appropriate {\em equilibrium} distribution function of the ions 
and electrons deviates respect to the usual Maxwell-Boltzmann (MB)
distribution, which is recovered in the weakly-interacting limit. These
small deviations are amplified by the barrier penetration mechanism yielding
non-negligible modifications of the solar neutrino fluxes~\cite{lis}.

Nuclear reaction rates in the solar interior (SI), which has a plasma parameter
$\Gamma \approx 0.1$, can be enhanced or suppressed 
depending on the
behavior of the high-energy tail of the equilibrium distribution function
of such weakly non-ideal plasma. Similar effects are expected in other
environments characterized by plasma parameters 
$0.1\lsim\Gamma\lsim 1$,
{\em e.g.}, recombination and nuclear rates in brown dwarfs, Jupiter core
and stellar atmospheres.

In the SI the conditions required by the Debye-H\"uckel are only approximately
verified. The plasma frequency $\nu_{pl}$ is of the same order of the 
collisional frequency $\nu_c$: $\nu_{pl}\approx\nu_{c}\approx 3-6 \times 
10^{17}$ sec$^{-1}$ and the screening radius is of the order of the
interparticle distance. In addition to many-body collisional effects,
electric microfields are present. The distribution
and the fluctuations of these microfields must be carefully considered,
since it modifies the usual Boltzmann kinetics.

In the SI all these effects that can modify the Boltzmann kinetics
are not very strong, contrary to the case of strong interacting Coulomb
plasmas ($\Gamma>1$)~\cite{yan}. 
Therefore, corrections to the usual MB distribution
are small and affect mainly the more energetic particles. Physical
quantities that depend on integrals over the entire distribution are almost
unaffected. However, nuclear reactions are strongly affected, when they
receive the main contribution from the high-energy tail of the
distribution, because of the Coulomb barrier.

In general, small deviations from the MB equilibrium distribution can
always be asymptotically described by~\cite{qua}
\begin{equation}
f(E)\approx \exp{\left[
     -\frac{E}{kT} - \hat{\delta} \left(\frac{E}{kT}\right)^2 
                 \right]}\, ,
\label{tsadi}
\end{equation}
where $\hat{\delta}$ is the Clayton parameter~\cite{cla}. In particular, it
has been shown~\cite{qua,lis} that, if one expresses the small corrections
to the usual kinetics by a suitable power expansion of the diffusion and
drift coefficients in the Fokker-Plank equation, the leading modification
of the distribution is indeed given by Eq.~(\ref{tsadi}).

In a more general context, when deviations are not necessary small,
it is necessary to generalize the Boltzmann-Gibbs statistics and to include
the possibility of non-extensive thermostatistics. Tsallis has recently
developed a statistical framework that naturally includes long-range
interactions and non-Markovian memory effects~\cite{ante}.
The particle velocity distribution predicted by this theory is
\begin{equation}
\label{tsdis}
f(v) = \left[ 1 + (q-1)\frac{m v^2}{2 kT}
      \right]^{1/(1-q)}
       \Theta\left[1 + (q-1)\frac{m v^2}{2 kT}\right]\, ,
\end{equation}
where $\Theta$ is the Heaviside step-function,
and the parameter $q$ can be connected by expansion to $\hat{\delta}$:
$q=1-2\hat{\delta}$. In the limit $q\rightarrow 1$
the Boltzmann-Gibbs statistics is recovered.

A specific physical model for the microscopic processes in SI that produces
the above equilibrium distribution is outlined in the following.

Each particle is affected by the total electric field distribution due to the
other charges in the plasma. The random electric microfields
are often expressed in term of the dimensionless parameter $F$
as $\langle {\cal E}^2 \rangle = (F \, e/a^2)^2$, where $4\pi a^3/3 = 1/n$;
the distribution of $F$ in plasmas depends
on the value of $\Gamma$~\cite{hoo,igle,roma,val,ebeling}.
These microfields have in general long time correlations, and can
generate anomalous diffusion.

The total microfield can be decomposed in three main components.
(i) A slow-varying (relative to the collision time) component due to the
collective plasma oscillations, which the particle sees as an almost
constant external mean field ${\cal E}$ over several collisions.
(ii) A fast random component due to particles within a few Debye radii,
whose effect can be described by an elastic diffusive cross section
$\sigma_1\sim v^{-1}$. When only this cross section is present, the
distribution remains Maxwellian even in presence of the slow mean field
${\cal E}$.
(iii) A short-range two-body strong Coulomb effective interaction, that
can be described by the ion sphere model~\cite{yan}. The strict enforcement
of this model as implemented by Ichimaru~\cite{yan} yields the elastic
cross section $\sigma_0=2\pi\alpha^2a^2$, where $a$ is the interparticle
distance, $\alpha$ an adimensional parameter whose order of magnitude can
be inferred from the parameter $F$ that characterizes
the microfields, $F\approx \alpha^{-2}$. 
Since $F^2 \sim 3/\Gamma\approx 40$, for $\Gamma = 0.07$, one can estimate
$0.4<\alpha<1$. In the present model, it is this component of the electric
field that turns out to be mainly responsible of the correction factor
$\exp{[-\hat{\delta} (E/kT)^2]}$.

In this framework, the stationary solution of the kinetic equation valid for
small deviations from the MB distribution can be shown to be:
\begin{equation}
f(E) \sim \exp{\left[ - \hat{\varphi} \frac{E}{kT}
              - \hat{\delta}  \left(\frac{E}{k T}\right)^2 
               \right]}  \, ,
\end{equation}
where 
\begin{equation}
\hat{\varphi}=\frac{\varphi}{1+\varphi} \ \ \ \ \ \varphi=\frac{9}{2} \kappa
\left(\frac{n k T}{Z e {\cal E}}\right)^2 \langle\sigma_1^2\rangle \, , 
\nonumber
\end{equation}
\begin{equation}
\hat{\delta}=\left (\frac{3 \langle\sigma_1^2\rangle}{\sigma_0^2} + 
\frac{1}{\delta}\right )^{-1} \ \ \ \ \ \delta=\varphi
\frac{\sigma_0^2}{3 \langle\sigma_1^2\rangle} \, , \nonumber 
\end{equation}
$\kappa=2 \, m_a m_b/(m_a+m_b)^2$ is the elastic energy-transfer
coefficient between two particles $a$ and $b$, and
$\delta$ is proportional to the square of the ratio of the energy densities
of the electric field and of the thermal motion.

In the small correction limit, relevant to the SI,
$\hat{\varphi} = 1 $ and the Clayton parameter 
is: 
\begin{equation}
\label{deltamodel}
\vert\hat{\delta}\vert \approx 
\frac{\sigma_0^2}{3 \langle\sigma_1^2\rangle}=12 \, \alpha^4 
\, \Gamma^2 \ll 1 \, .
\end{equation}
While we presented only a specific model, it is already sufficient to
introduce the kind of mechanisms that modify the MB distribution.
It is especially important
to study the time and spatial correlation of the microfields, and how the
physical phenomena of interest, {\em e.g.}, subbarrier nuclear reactions,
generate the appropriate scale separation of the random field
contributions.

The presence of even a tiny deviation from MB in the solar core produces
large changes of the subbarrier nuclear reaction rates and, consequently,
of the predicted neutrino fluxes. Following the general homology 
relationships for the variations of physical inputs,
see for instance Ref.~\cite{caste}, we can estimate the effect of
the non-Maxwel\-lian distribution on the fluxes~\cite{lis}:
\begin{equation}
R_j=\frac{\Phi_j}{\Phi_j^{(0)}}=e^{-\hat{\delta}_j \beta_j} \, ,
\label{flussi}
\end{equation}
for the fluxes $j=$ $^7$Be, $^8$B, $^{13}$N and $^{15}$O, while we
use the solar luminosity constraint to determine the $pp$ flux, 
$R_{pp} = 1+0.087\times(1-R_{Be})
                       +0.010\times(1-R_{N})
                       +0.009\times(1-R_{O})$, and keep fixed
the ratio $\xi\equiv\Phi_{pep}/\Phi_{pp}=2.36\times 10^{-3}$. 
The power index $\beta_j$, that compares in Eq.~(\ref{flussi}), depends
from the  nuclear reaction considered and its value has been taken from
Ref.~\cite{caste}. 
In principle, there could be a different parameter $\hat{\delta}_j$ for
each reaction and it should be possible to calculated them from the 
specific interactions in the solar plasma core. For the purpose of
estimating the possible effects of this mechanism on the solar neutrino
fluxes, we used two simple models where $\delta_j$ were used as free
parameters. 

In the first model, we have used the same $\hat{\delta}$ for all the
reactions; the best fit to the experimental data give $\hat{\delta}=0.005$
with a corresponding $\chi^2=35$. In the second model, we have fitted
two different $\hat{\delta}$'s, one ($\hat{\delta}_{(17)}$) for the the
$p+{}^7$Be reaction and the other ($\hat{\delta}_{(34)}$) for the
${}^3$He + ${}^4$He reaction. The best result gives $\chi^2=20$ with
$\hat{\delta}_{(17)}=-0.018$ and $\hat{\delta}_{(34)}=0.030$.
The negative value of $\hat{\delta}_{(17)}$ means that the corresponding
distribution has an enhanced tail and that the $p+{}^7$Be reaction
rate increases. 

In spite of the fact that the values of $\chi^2$ are much smaller than
the ones in the SSM ($\chi^2_{SSM}>74$), they are still large: this
mechanism cannot solve the solar neutrino problem. The best values of
$\hat{\delta}$ are of the order of the ones found using Eq.~(\ref{deltamodel})
with $\alpha \approx 0.4$ and $\Gamma \approx 0.07$, and they have been
found not incompatible with the constraints from helioseismic
observations~\cite{scilla}. Even if $\hat{\delta}$ is small, it has
non-trivial consequences on the neutrino fluxes: the boron (beryllium)
flux can change of as much as 50\% (30\%). 

In conclusion we have shown that it is plausible that the equilibrium
velocity distributions in SI are not Maxwellian and they could
follow from Tsallis statistics. The parameter $\hat{\delta}$ that characterizes
the generalized distribution is related to the microscopic collective
processes in the plasma. In particular, it is of great interest to consider
the influence of the random electric microfields on the diffusion processes,
and the possibility of characterizing their effects by average quantities
(mean fields and cross sections) over different scales. In the future, one
needs detailed studies of the time (memory effects) and spatial
correlations of these microfields.
Although these effects cannot by themselves solve the solar neutrino
problem, they can produce large changes of the predicted fluxes and should
be taken into account when looking for other solutions to the solar
neutrino problem.

\end{document}